\documentclass[a4paper,floatfix,rmp,superscriptaddress]{revtex4}
	\usepackage{amsmath,amssymb,amsfonts}    
	\usepackage[latin1]{inputenc}
	\usepackage{graphicx}
	\usepackage{url}
\begin{document}

	\title{Ambiguity in language networks}

  \providecommand{\ICREA}{ICREA-Complex Systems Lab, Universitat Pompeu Fabra (GRIB), Dr Aiguader
													80, 08003 Barcelona, Spain}
	\providecommand{\IBE}{Institut de Biologia Evolutiva, CSIC-UPF, Pg Maritim de la Barceloneta 37,
												08003 Barcelona, Spain}
	\providecommand{\SFI}{Santa Fe  Institute, 1399 Hyde  Park Road, Santa Fe NM 87501, USA}

	\author{Ricard V. Sol\'e\footnote{Corresponding author}}
		\affiliation{\ICREA}
		\affiliation{\IBE}
		\affiliation{\SFI} 
	\author{Lu\'is F. Seoane}
		\affiliation{\ICREA}   
		\affiliation{\IBE}

	\vspace{0.4 cm}
	\begin{abstract}

		\vspace{0.2 cm} 

    Human language defines the most complex outcomes of evolution. The emergence of such an
elaborated form of communication allowed humans to create extremely structured societies and manage
symbols at different levels including, among others, semantics. All linguistic levels have to deal
with an astronomic combinatorial potential that stems from the recursive nature of languages. This
recursiveness is indeed a key defining trait. However, not all words are equally combined nor
frequent. In breaking the symmetry between less and more often used and between less and more
meaning-bearing units, universal scaling laws arise. Such laws, common to all human languages,
appear on different stages from word inventories to networks of interacting words. Among these
seemingly universal traits exhibited by language networks, ambiguity appears to be a specially
relevant component. Ambiguity is avoided in most computational approaches to language processing,
and yet it seems to be a crucial element of language architecture. Here we review the evidence both
from language network architecture and from theoretical reasonings based on a least effort argument.
Ambiguity is shown to play an essential role in providing a source of language efficiency, and is
likely to be an inevitable byproduct of network growth.

	\end{abstract}


	\maketitle

	\tableofcontents{}

	\section{Introduction}
		\label{sec:01}

    One of the latest and yet more profound evolutionary transitions involved the appearance of a
new form of communication. Human language represented the triumph of non-genetic information, in a
scale and quality that allowed a virtually infinite repertoire of meaningful constructs out of a
collection of basic lexical units. Cultural evolution became a major player in shaping the character
of human societies \cite{MaynardSzathmary1995, HauserFitch2002}.

    It is fair to say that language, and human language in particular, has received the most
dedicated multidisciplinary efforts. These include a vast range of fields, from genetics and
anthropology to cognitive sciences, artificial intelligence or game theory. And yet, despite its
undeniable importance, the origins of language remain largely unknown. Moreover, a graded transition
to this complex form of communication does not exist. It is a sharp, drastic change what mediates
between human languages and other animal communication systems. This enormous gap makes difficult to
retrieve information by comparing our tongues to any midway stages.

    We deal with a complex system that involves multiple scales and intricate interactions between
levels and component units \cite{AltmannEsposti2012}. As such, a proper approach to its complexity
needs a framework that explicitly considers systemic properties. Born by this complexity, language
displays all kinds of apparently odd features, from the sometimes quirky appearance of syntactic
rules to the ubiquitous presence of ambiguity. Ambiguity is specially puzzling: it seems to make
little sense when we consider language from an engineering perspective or even under a standard
optimization view based on communicative pressures \cite{PinkerBloom1990, Chomsky2002}. Under this 
view, selection for comprehensible symbols would act removing unreliable components, thus 
reducing ambiguous features to the minimum. 
 
    Following the optimization line of thought, the ultimate basis of our discourse will be that a
{\em least effort} principle is a driving force of languages. Always focused on this argument, in
this paper we present recent theoretical advances that share a common systems-level perspective of
language structure and function. We adopt a non-reductionist approach towards human language
\cite{Kaufman1993, SoleGoodwin2001, Ke2004} that largely relies on a network view of its
structure -- closer to a structuralist view of evolution. Within this view, constraints and genuine,
endogenous features manifest themselves promoting (and being masked behind) universal statistical
regularities. The discussed theoretical arguments are preceded by the description and discussion of
experimental facts -- always following the same systemic approach -- that clearly show the kind of
universal traits that we refer to and that happen to pervade every known language.\\

    After discussing some striking empirical universal regularities of human language in sections
\ref{sec:02} and \ref{sec:03} and their connections to ambiguity in section \ref{sec:04} we briefly
present experimental support of the least effort argument and analyze in detail some of its
theoretical consequences in sections \ref{sec:05} and \ref{sec:06}. We will also see how some of
these consequences link back to the ever present statistical regularities mentioned earlier.
Finally, in section \ref{sec:07} we sketch out open questions and research lines that could further
our understanding about the fascinating matter of human language.

	\section{Scaling in language}
		\label{sec:02}

    Language structure has been very often contemplated under the perspective of word inventories.
The properties of isolated words and how these properties can be used to classify them within given
general groups provide a first way of studying language architecture. The abundance of words, how
they become adopted over language acquisition, or how different levels of language structure shape
their relative importance define major research areas within linguistics. When exploring word
inventories, one is faced with a dual character of languages that confronts the heterogeneity of
tongues with the deep universality of a variety of their traits.

    So, on the one hand languages are diverse. This is reflected in several features displayed by
its constituents. Word inventories obviously differ from one dialect to another. Many
characteristics, such as the number of letters in a word -- for example, show a statistical pattern
with a distinctive single-hump distribution, but the average number of letters is rather different
across languages. In Mongolian or German this is close to 12 letters per word, whereas for Croatian
or Serbian this drops down to around seven. The diversity in
this trait might originate in historic contingencies idiosyncratic of each language and is not -- a
priori -- the kind of universalities that we wish to study.

    On the other hand, it has been shown that all languages seem to share some remarkable universal
patterns, best exemplified by the so called Zipf's law \cite{Zipf1949}. Earlier noted by other
authors, but popularized by G. K. Zipf, this law states that the frequency of words in a given word
inventory -- such as the one we can obtain from a book -- follows a universal power law.
Specifically, if we rank all the occurrences of words in a given text from the most to the less
common one, Zipf's law states that the probability $p(s_i)$ that in a random trial we find the
$i$-th most common word $s_i$ (with $i=1,...,n$) falls off as:
			\begin{eqnarray}
				p(s_i) &=& \frac{1}{Z}i^{-\gamma}, 
				\label{eq:02.01}
			\end{eqnarray}
with $\gamma \approx 1$ and $Z$ the normalization constant -- i.e., the sum $Z=\sum_{i \leq n}
i^{-\gamma}$. We can observe this regularity in any modern human language when analyzing any
adequate corpus. This is the kind of traits that we are interested in, and of which we demand an
explanation with the hope of gaining a deeper understanding about the origins of language or the
constrains that shape it. \\

			\begin{figure*}
				{\centering 
					\includegraphics[width=15cm]{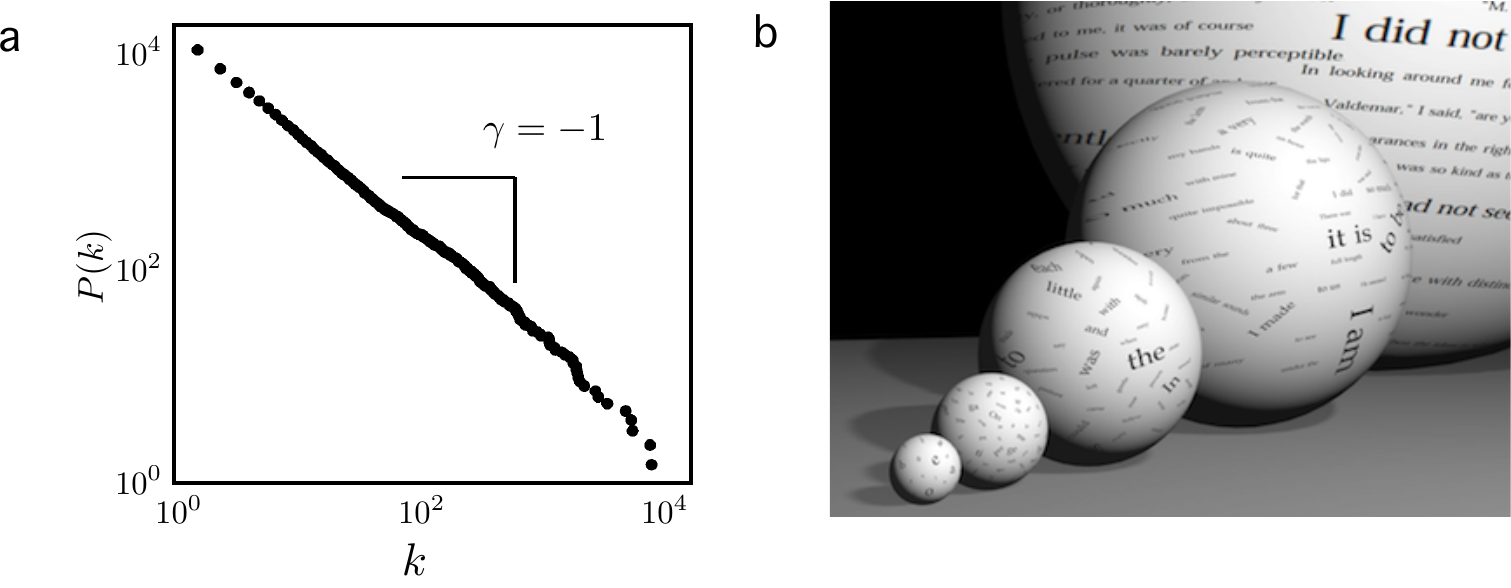}

    \caption{A seemingly universal feature of all known human languages is Zipf's law,
illustrated in figure (a) from the rank-abundance statistics obtained using Melville's Moby Dick
(see text). Moreover (b) language contains multiple levels of nested complexity, illustrated here by
means of an idealized collection of spheres whose size rapidly grows as the objects being considered
at one level are combined to obtain those in the next level. Letters and syllabus are the first
levels, followed by words and pairs of words and eventually to sentences. The diagram actually
underestimates the real proportions of combinatorics. }

					\label{fig:02.01}
				}
			\end{figure*}

    Roughly speaking, Zipf's law tells us that the most frequent word will appear twice as often as
the second most frequent word, three times as often as the third one, and so on. Instead of using a
word's rank, an alternative form considers the use of the standard probability $p(m)$ that we come
across a word that is repeated $m$ times throughout a text. Then the corresponding Zipf's law scales
as:
			\begin{eqnarray}
				p(m) &=& \frac{1}{Z} m^{-\alpha}
				\label{eq:02.02}
			\end{eqnarray}
where now the normalization constant is $Z=\sum_{m \leq M} m^{-\alpha}$ with $M$ the maximum
observed frequency. Now the scaling exponent is $\alpha = 2$. In figure \ref{fig:02.01}{\bf a} the
frequency-rank distribution of words collected from Herman Melville's Moby Dick is shown in
logarithmic scale. The scaling law $P(k) = k^{-\gamma}/Z$ is plotted against the rank $k$. The
logarithmic plot provides a direct way of testing the presence of a scaling law, since it gives a
linear relationship:
			\begin{eqnarray}
				\log p(k) &=& \log \left [ \frac{1}{Z} k^{-\gamma} \right ] \nonumber \\
					&=& \log \left[ \frac{1}{Z} \right] - \gamma \log k, 
				\label{eq:02.03}
			\end{eqnarray}
the slope of which is the scaling exponent $\gamma$. 

    The widespread, virtually universal presence of Zipf's law in all known languages, and perhaps
even in the context of DNA and the genetic code \cite{MantegnaStanley1994, Searls2002,
ObstProkopenko2011} suggests two potential interpretations. It might be the case that the observed
scaling is so widespread that it is essentially a meaningless signal. (Note the discussion about
Zipf's law in random texts \cite{FerrerSole2001c}.) The other possibility is that its universal
presence has to do with some relevant feature shared by all languages, perhaps associated to some
deep functional role. Given the disparate trajectories followed by human languages over their
evolution, it seems unlikely that such a unique scaling law would be so robust unless it involves a
relevant constraint.\\

    An additional component related to the logical organisation of language deals with its enormous
combinatorial potential. Language defines a non-genetic form of heredity and as such allows rapid
cultural exchanges, the formation of a collective memory, and an enormous plasticity while facing
environmental challenges. Its success is tied to the brain's capacity for storing a large number of
communication elements. However, an inventory of words can only be part of the whole story. Another
important aspect must be the associations that these units can build between them and that will be
treated in more detail in the next section. Let us explore first the scaling facet of such
associativity to have a scope of the relevance of the generative power of language.

    Words are combined and related to each other in multiple ways. Such combinatorial potential
pervades all linguistic levels from phonemes to whole texts. As we move towards higher levels, the
potential universe of objects expands super-exponentially (figure \ref{fig:02.01}{\bf b}). We can
appreciate this inflationary behavior explicitly when moving from words to sentences to texts. Let
us assume a set of words  ${\cal L'} $ is sampled from the  whole repertoire of words defining a
language ${\cal L}$ (i. e. ${\cal L'} \subset {\cal L}$).  Our set ${\cal L'}$ is finite and
involves $\vert {\cal L'} \vert = N_w$ words. Of course the combinatorial nature of word
arrangements easily explodes with $N_w$. Now consider a finite (but long) written text,  to be
indicated as $\cal T$.  It is composed by a set of $M$ sentences $S_{\mu}$, each one formed by an
ordered, sequential collection of words extracted from $\cal L'$:
			\begin{eqnarray}
				S_{\mu} &=& \{ w_{1,\mu}, w_{2,\mu}, ..., w_{n_{\mu},\mu}\}
				\label{eq:02.04}
			\end{eqnarray}
with $\mu=1,2....M$ and thus we have our text defined as the union:
			\begin{eqnarray}
			  {\cal T} &=& \bigcup_{\mu=1}^M S_{\mu}
			  \label{eq:02.05}
			\end{eqnarray} 
If we indicate by $\vert S_{\mu} \vert$ the length of a given sentence, the average sentence size in
$ {\cal T}$ will be
			\begin{eqnarray}
				\langle S \rangle &=&  {1 \over M} \sum_{\mu} \vert S_{\mu} \vert 
				\label{eq:02.06}
			\end{eqnarray} 
A very rough first approximation assuming that all components can be combined in similar ways --
i.e. leaving syntactic constrains aside --  provides a total number of (possible) sentences as given
by the power law:
			\begin{eqnarray}
			  \vert {\cal T} \vert &\sim& N_w^{\langle S \rangle}
			  \label{eq:02.07}
			\end{eqnarray}		
which gives, for $N_w \approx 80000$ and $\langle S \rangle \approx 7$ (two reasonable estimates) a
hyperastronomic number: $2.097 \times 10^{34}$. In natural language, many of these combinations will
never appear, most of words will be extremely rare and a few of them extremely frequent (as we saw
above) since there exist nontrivial rules for a string of symbols make sense as a word of ${\cal
L}$. The plausibility of a sentence existence and its frequency will be constrained as well because
there are further nontrivial (syntactic) rules for the use of words from ${\cal L}$ in a real
context. Nevertheless, this quick calculation allows us to grasp the scope of the expressive power
of this system.

    The enormous potential for combination that is present in human language embodies the uniqueness
of such complex form of communication. No other species in our planet shares such a spectacular
capacity and a chasm seems to exist between us and all the other species inhabiting our planet. This
uniqueness is also interesting for another reason. Major innovations that have occurred through
evolution have been found independently a number of times. Multicellularity, sight or sex have
emerged in many different groups through different paths \cite{MaynardSzathmary1995, Gregory2008} 
thus indicating that the same basic innovations can be obtained following different paths.  
By contrast, our complex communication system that we use as a species, is unique
\cite{MaynardSzathmary1995}. No othe rparallel experiments in evolution leading to such achievement 
have taken place.

    However, storing words is one thing; combining them, another; and being able to relate each
other in a flexible, efficient manner is yet another one. Our potential for storing a large
inventory of words together with an astonishing potential of relating them in complex ways through
intricate paths (sentences being just one of them) is at the core of the evolutionary success of
humans. In this paper we consider language organization in terms of a statistical physics picture,
where networks instead of word inventories play a central role. By using them, we will argue that
ambiguity is an expected feature of human language, and a specially relevant and perhaps inevitable
one. It is ambiguity what hides behind Zipf's law and an essential element that makes our use of
language so efficient and flexible.

	\section{Small world language networks}
		\label{sec:03}

    In our previous illustration of the combinatorial potential of language, we used sentences as
higher-order structures obtained as linear chains that combine words in syntactically meaningful
ways. Sentences provide us with a first example for the kind of recursive linguistic structures that
we are capable of forming. They will also serve us to introduce networks and how language can be
interpreted in terms of these complex webs.

    The simplest case of language network that can be introduced is defined in terms of 
co-occurrence \cite{FerrerSole2001a}. Two words in a sentence that appear one after the other are
said to co-occur. We will build a graph (a network) using these words and their co-occurrence as
follows: Words $w_i$ $(i=1, ..., N_w)$ are the fundamental units, defining a set $W$. The
relationships between words are encoded in a matrix $\Gamma=\{a_{ij}\}$ called the {\em adjacency
matrix}. An undirected link $a_{ij} = 1 = a_{ji}$ will be defined between two words $w_i, w_j \in W$
if they follow one another within at least one sentence (otherwise the matrix element is set to
$a_{ij}=0=a_{ji}$). The resulting {\em language production network} (LPN) $\Omega_L$ is thus defined
as a pair $\Omega_L=(W,\Gamma)$, where $\Gamma=\{a_{ij}\}$ constitutes the set of unweighted links
of the graph. It should be noticed that the mapping $\Gamma:W \rightarrow W$ is expected to capture
some of the underlying rules of word ordering. This web provides in fact a glimpse to the production
capacity of the underlying grammar structure and shares, as we will see below, a large number of
common traits with syntactic webs \cite{FerrerKohler2004}.

			\begin{figure*}[htb]
				{\centering 

				\includegraphics[width=12 cm]{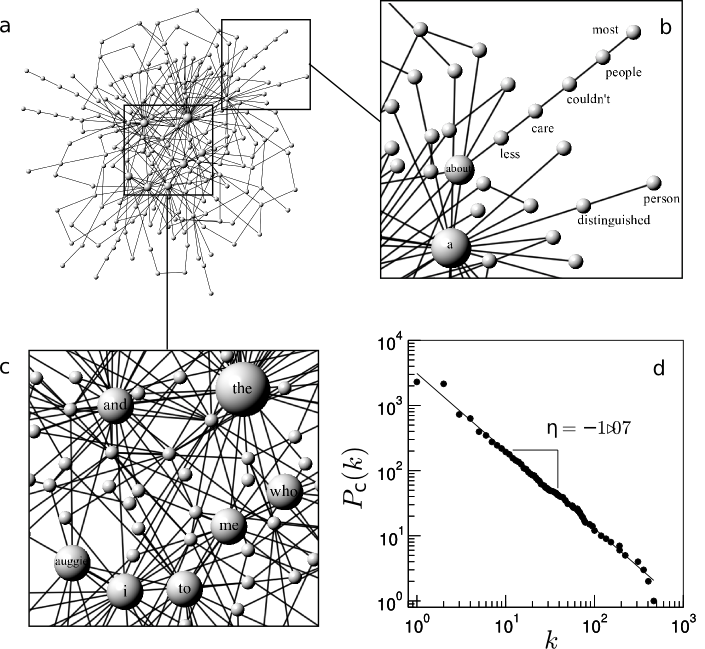}

    \caption{A language network can be build in different ways. The simplest one is considering 
co-occurrence between words within sentences from written corpuses. Here (a) we have used the first
chapter of Paul Auster's ``Augie Wren Xmas Tale", from which we draw our network. Each ball is a
different word, whereas an undirected link between two balls indicates that those words appeared one
after the other within a sentence in the text. Two parts of the web are zoomed in (b) and (c). In
(b) we can see that some words have a very large number of links with others and are referred to as
``hubs", whereas most words have just one or two connections. In (c) we also see multiple linear
structures and chains associated to particular sentences. LPNs follow scale-free degree
distributions, as exemplified in (d). This is the same statistical feature of word frequency
illustrated by the Moby Dick data set (see sect. \ref{sec:02} and fig. \ref{fig:02.01}{\bf a}).}

					\label{fig:03.01}
				}
			\end{figure*}

    In figure \ref{fig:03.01}{\bf a} we display an example of LPN network. This particular one has
been obtained from the words that appear in Paul Auster's short story {\em Augie Wren Xmas tale}.
Here spheres correspond to specific words and connections among them indicate that the pair of word
co-occurred at least within one sentence throughout the tale. The size of the spheres has been
increased in some cases to indicate their high frequency of appearance in the text. Several
interesting features need to be noticed. One is that the network is highly heterogeneous: a vast
majority of words have only one or two links with others, whereas a small number of them (the hubs)
have a very large number of connections. These super connectors can be seen in figure
\ref{fig:03.01}{\bf c} and correspond to words that are very common and highly ambiguous. Figure
\ref{fig:03.01}{\bf b} gives us a glance of the ``local" organization stemming from the sentence
structure. We can actually read well defined chains that make sense in a given direction. These
readable chains become less and less common as the size of the word inventory grows and more and
more crossings occur.

    A distribution of connections, or {\em degree distribution} $P(k)$, can be defined by measuring
the number of links $k$ of each node (also known as its degree) and calculating the relative
frequencies for each $k$. In a randomly wired graph of $N$ nodes, where we simply connect every two
elements with some probability $p$ the number of links associated to a randomly chosen word would
follow a Gaussian distribution, centered around the average degree value $\langle k \rangle =
p(N-1)/2$. We call such a graph {\em homogeneous} because the average value represents fairly well
everything that can be awaited of the graph. But many real networks -- including language graphs --
follow a functional form that displays a scaling law, namely
			\begin{eqnarray}
				P(k) &=& \frac{1}{Z} k^{-\alpha}. 
			  \label{eq:03.01}
			\end{eqnarray}
Once again, we have $Z = \sum_k k^{-\alpha}$ and, for all LPN networks, $\alpha \approx 2$. Let us
note once more the remarkable universality of this observation: for any language, from any adequate
collection of sentences, despite the disparity that both elements (languages and sentences -- and
collections of sentences, indeed) can present we will derive such a degree distribution with roughly
the same exponent $\alpha$; just as if some inner mechanisms of the human language were eventually
responsible of such scaling. As opposed to the Gaussian, these kind of power law distributions
feature an extreme variability that the average alone cannot capture. This is a consequence of the
existence of a miscellany of structures within the network. The real world example from Auster's
short story is shown in figure \ref{fig:03.01}{\bf d}, where we have used  (to smooth out the
statistics) the cumulative distribution, defined as
			\begin{eqnarray}
				P_>(k) &=& \sum_k^M P(k) \sim \int_k^M P(k) dk \sim k^{-\alpha+1} = k^{-\gamma}. 
				\label{eq:03.02}
			\end{eqnarray}
We find an exponent $\alpha \sim 2$, which is actually the same that we observed in Zipf's law (in
its frequency form). This is not surprising, since there is an almost perfect correlation between
the frequency of a given word and the number of co-occurrences it can establish within $W$.
Therefore, it could be argued that the only thing that matters is the frequency distribution of
words: this would eventually determine the degree distribution. However, there is more to the
structure of the network than this power law distribution of its degree $k$. To appreciate it we
must look at some other traits.

    A randomly connected graph following the previous $P(k)$ scaling would not recover many
observable properties exhibited by the original graph based on co-occurrence. As an example, there
is a widespread feature that is present in the LPN and not in a randomized version of it: hubs are
usually not connected in the former but they can be so in the later. This particular result tells us
that, despite not being a true syntactic network, LPNs do preserve some essential constraints
associated to syntactic rules.

    There is another interesting property. The LPN graph is sparse: the average number of
connections per word is small. Despite this sparseness and the local organization suggested by the
previous features, the network is extremely well connected. In complex networks theory, this is
known as a {\em small world} graph \cite{WattsStrogatz1998, AlbertBarabasi2002}. Small world
networks were first analyzed by Stanley Milgram in the context of social ties within a country
\cite{Milgram1967}. It was found that only a small number of links separates, within the network of
social acquaintances, two randomly chosen individuals. Since a given country involves millions of
humans, the basic result -- that only about six jumps are needed (on average) to connect any two
random persons -- was highly surprising. This qualitative property can be quantified by means of the
{\em average path length} ($D$) defined as $D = \langle D_{min}(i,j) \rangle$ over all pairs $w_i,
w_j \in W$, where $D_{min}(i,j)$ indicates the length of the shortest path between two nodes. Within
the context of a LPN, a short path length means that it is easy to reach a given word $w_i \in W$
starting from another arbitrary word $w_j \in W$. The path cannot be interpreted here in terms of
meaningful trajectories (such as sentences) but instead as a measure of accessibility.

    An additional measure of network organization that characterizes small world graphs is the so
called {\em clustering coefficient} ($C$). It is defined as the probability that two vertices
(words, in out context) that are neighbors of a given vertex are neighbors of each other as well. In
order to compute the clustering, we associate to each word $w_i$ a neighborhood $\Gamma_i$, defined
as the set of words linked to $w_i$, i. e.
			\begin{eqnarray}
				\Gamma_i = \{ w_k \in W \; \vert \; a_{ik}=1 \} 
				\label{eq:03.03}
			\end{eqnarray}
Each word $w_j \in \Gamma_i$ has co-occurred at least once with $w_i$ in some sentence. The words in
$\Gamma_i$ can also be linked among them. The clustering $C(\Gamma_i)$ of this set is defined as the
fraction of triangles found, compared to the maximal number expected from an all-connected scenario.
Formally, it is given by:
			\begin{eqnarray}
				C(\Gamma_i) = {1 \over k_i(k_i-1)}\sum_j \sum_{k\in\Gamma_i} a_{jk} 
				\label{eq:03.04}
			\end{eqnarray} 
and the average clustering is simply $C=\left<C(\Gamma_i)\right >$. Many triangles in a sparse graph
indicate an excess in local richness of connections. Such an excess needs to be compared with a null
model of random connections among words -- i.e. with a randomized version of the LPN as we did to
compare the likelihood that the hubs are connected.

    Concerning the average path length, for random graphs with Poissonian structure -- i.e. with
nodes simply connected with a probability $p$ and thus their degree distribution following the
rather unremarkable Gaussian distribution -- it is possible to show that we have a logarithmic
growth in the number of degrees of separation with $N$ \cite{WattsStrogatz1998, AlbertBarabasi2002}:
			\begin{eqnarray}
				D_{random} &\approx& {\log N \over \log \left<  k  \right>};
				\label{eq:03.05}
			\end{eqnarray} 
whereas the clustering is expected to decay inversely with system size -- i. e. 
			\begin{eqnarray}
				C_{random} &\approx& {1 \over N}. 
				\label{eq:03.06}
			\end{eqnarray} 		

		On a first approximation, it is said that a network is a {\em small-world} when $D \approx
D_{random}$ whereas the clustering coefficient is much larger $C\gg C_{random}$
\cite{WattsStrogatz1998, AlbertBarabasi2002}. LPNs happen to be small worlds, as remarked above.
This nature of LPNs and other language networks tells us that despite their locally ordered,
correlated structure (far from that of a random graph) association and routing between words can be
highly efficient.

    Network theory does not offer a full explanation for the cognitive substrate responsible for
word association and optimal search -- this last property being related to the easy navigation that
small worlds enable. This theory does provide, though, a valid formal framework within which
relevant questions can be consistently stated. Hopefully, the answers attained also constitute
compelling knowledge about human language.

	\section{Ambiguity in semantic networks}
		\label{sec:04}

    The relational nature of language can be analyzed from different scopes. They include semantics,
syntax, morphology and phonology \cite{Pustejovsky1991, Pustejovsky1995, Chomsky2000, Scalise1984,
Trubetskoi1939}. They define the different relationships between units and the structures made by
such units. We saw a syntactic example in the previous section. Moreover, at the community level
social interactions also describe a web within which languages are enforced. This social structure
can play a determinant role, for example, in the success or failure of a contingent linguistic trait
and even in the emergence of further universal regularities \cite{SoleFortuny2010}. We see that
network theory is not only useful but perhaps inescapable to understand our communication system.
All these networks must somehow contain information concerning the way in which components -- {\em
generally, but not necessarily, words} -- are organized within sentences or how they are related in
terms of their semantic content. The links can thus have a very different nature in each graph and
the overall patterns of organization of such graphs do not need to be the same.

    A prominent subfield of linguistics, semantics has been traditionally defined as the study of
the meaning of (parts of) words, phrases, sentences, and texts. Semantic organization is a widely
explored topic in psycholinguistics. As a search for an adequate characterization of meaning,
semantic relations have strong ties with memory and categorization. Semantic relations are also
known to deteriorate in patients with Alzheimer's disease and other types of brain impairment
\cite{ChanSalmon1997}. Such a semantic decline can also be appreciated in the kind of properties
(e.g. Zipf's law) that we are interested for other diseased patients \cite{Ferrer2005a}.\\

			\begin{figure*}[htb]
				{\centering 

				\includegraphics[width=11 cm]{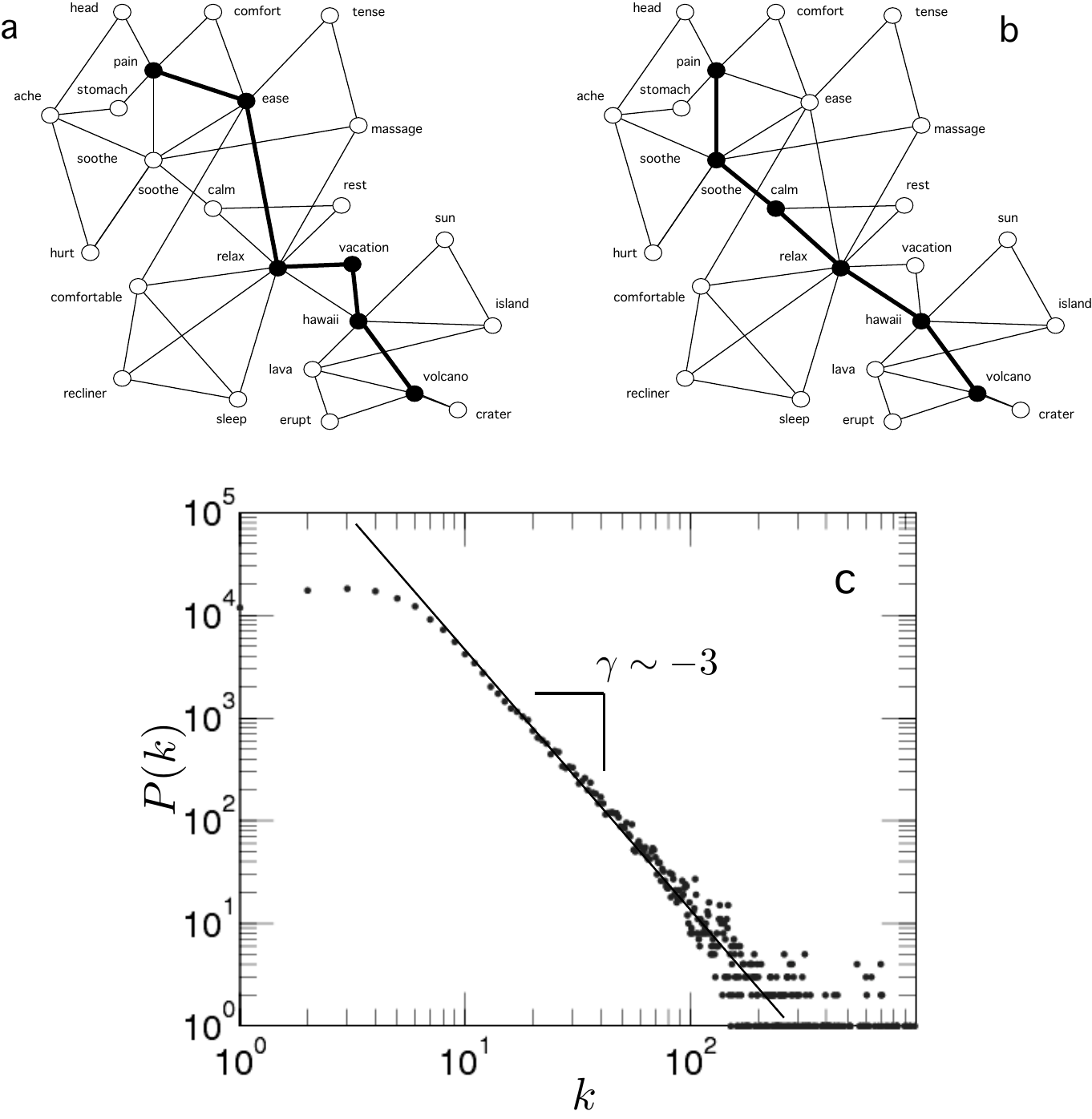}

        \caption{A simple network of semantic relations among lexicalised concepts. Nodes are
concepts and links, semantic relations between concepts. This would correspond to a very small
subset of a vast set of words and semantic relationships. Associations between words allow us to
navigate the network. Locally, the number of triangles is very large, allowing multiple ties
among semantically related words -- and contributing to a high clustering, as seen in the text.
Moreover, given two words, such as ``volcano" and ``pain" can be linked through different paths, two
of which are illustrated here using thick lines. The degree distributions associated to these
semantic graphs are broad, with fat tails. In (c) we display the distribution of links for WordNet,
with a scaling exponent close to three.}

					\label{fig:04.01}
				}
			\end{figure*}

    Semantic networks can be built starting from individual words that lexicalise concepts and by
then mapping out basic semantic relations such as isa-relations, part-whole, or binary opposition.
They can potentially be built automatically from corpus data \cite{KinouchiRisau2002,
MotterDasgupta2002, SigmanCecchi2002, HolandaSeron2004, SteyversTenenbaum2005, GoniVilloslada2011}
and also from retrieve experiments in which subjects are asked to quickly list down words as they
come to their minds \cite{SteyversTenenbaum2005, GoniVilloslada2011}. One of the most interesting
efforts in understanding the organization of semantic relationships is the Wordnet project
\cite{Miller1995, Fellbaum1998}. This data set explicitly defines a graph structure where words from
the English lexicon are connected through various kinds of semantic links. A possible subset of such
kind of web is displayed in figure \ref{fig:04.01}{\bf a}-{\bf b}. As pointed out by Sigman and
Cecchi \cite{SigmanCecchi2002} mental concepts emerge as a consequence of their interrelationships,
and meanings are often related through chains of semantic relations. Linking ``stripes'' with
``lion'' requires following a mental path through a sequence of words, such as 
lion-feline-tiger-stripes \cite{SteyversTenenbaum2005}. Different paths are possible on a semantic
network -- as exemplified in figure \ref{fig:04.01}{\bf a}-{\bf b} -- and experience shows that we
find them easily despite the very large set of items potentially available.

    The efficient character of the semantic network is associated to an important, universal, and
yet apparently undesirable property of language: polysemy. All languages exhibit polysemy, meaning
that a given word form corresponds to two or more meanings. At first sight we would think that
polysemy is a rather undesirable feature, since some ideal language should be expected to avoid such
ambiguity. The analysis of the large-scale architecture of semantic networks reveals a likely reason
for polysemy to exist and be so widespread. The answer lies on the global organization of these
graphs which are both highly heterogeneous and exhibit the small world phenomenon. The network
analysis of Wordnet shows a scale-free structure (figure \ref{fig:04.01}{\bf c}) where most elements
would be more specialized, and thus semantically linked to just a few others. By contrast, a few of
them would have a large number of semantic links. As before, we have a degree distribution $P(k)
\sim k^{-\alpha}$, now with $\alpha \sim 3$ and thus a higher scaling exponent that indicates a much
faster decay in the frequency of high-degree elements. This network is a small world {\em provided
that polysemy is included}. The high clustering found in these webs favors search by association,
while the short paths separating two arbitrary items makes search very fast
\cite{MotterDasgupta2002} even if distant fields need to be reached. Additionally, as discussed in
\cite{SteyversTenenbaum2005}, the scale-free topology of semantic webs places some constraints on
how these webs (and others mentioned above) can be implemented in neural hardware. This is a
remarkable example of how statistical regularities could be hiding a very relevant constrain of
language evolution.

    To summarise, the mapping of language into networks captures novel features of language
complexity far beyond word inventories. It provides further evidence for universal traits shared by
all languages and how to characterise and measure them. More interestingly, they suggest novel ways
of approaching old questions related to language efficiency and how it might have evolved. But they
also allow us to formulate new questions that could not be expressed without using the network
formalism. Among them, how these network patterns might emerge and how they might be linked to
Zipf's law. In the next section, we will review a model of language evolution that also involves
graphs and that is based on an early proposal by Zipf himself. That model provides a first
approximation to the potential components that make human language unique. It turns out that
ambiguity might actually be a key component behind some of our more remarkable singularities.

	\section{The Least-Effort Language agenda}
		\label{sec:05}

     As we insisted throughout the text: statistic regularities are a narrow window that allows us
to glimpse the existence of universal laws driving the emergence and evolution of human languages.
Zipf's law remains the most singular of such universal observations. Opposed to partial collections
of words -- such as the analysis performed on {\em Moby Dick} in section \ref{sec:02} -- a careful
analysis of extensive corpora clearly indicates that the whole of a language does not feature the
pattern observed by Zipf \cite{FerrerSole2001b, PetersenPerc2012}. Just a {\em core vocabulary} does
so, but the observation remains universal anyway. Furthermore, recent analysis indicate that
diseased patients as well as lexicon not in the core might follow a version of Zipf's law with a
generalized exponent $\gamma \ne 1$ \cite{FerrerSole2001b, Ferrer2005a}. In sight of this evidence,
the general scientific intuition has a broad consensus about the importance of Zipf's law and
efforts to find model explanations to it do not diminish over time.

    In its original account, Zipf proposed that a tension between minimizing user's efforts and
maximizing the communication power of a language would be the main driver towards the statistic
regularity that he observed empirically, thus he coined the \emph{least effort language} principle
\cite{Zipf1949}. Our main concern in this section is not necessarily Zipf's law, but the least
effort optimization as a mechanistic driving force -- which, anyway, has been shown to be a
mechanism for the generation of scale-free distributions \cite{ValverdeSole2002}. There are strong
evolutionary reasons why a least effort principle might be acting upon human languages. To
appreciate the selection for least effort in communication we can adopt any of two complementary
view points -- both of which are visited in \cite{Wray2002}. On the one hand we could argue that a
human group with a more efficient code could enjoy an evolutionary advantage over other groups.
Those with less adequate dialects would be selected against and their tongues would perish with
them. The other possibility is to look at each language as a system enduring natural selection. We
can conceive different codes simultaneously spreading over a population. Those fitter to be
transmitted by humans -- i.e. those better coping with our biological, social, and technological
constrains -- would be naturally selected for and become dominant. Because the fitness now is the
ease of tongues to humans we can see a least effort driving language evolution quite directly, not
necessarily through an intermediate step of human selection.

    How can we approach language evolution from a sensible facet? There are in principle multiple
ways and scales of approximation that can be used. They span an enormous range of views, from 
game-theoretic models to computational linguistic or language evolution in embodied, robotic agents.
Perhaps the answer to previous questions needs to be tied to another, more basic one: What do we
want to understand? Here we are concerned with ambiguity as part of the fabric of language
organization. We would like to understand if ambiguity plays any role in how the previous scaling
laws emerge and why there might be sharply defined classes of languages -- perhaps separated by some
sort of barrier -- thus directly tackling the harsh gap between human and any other form of
communication. Following the steps indicated in \cite{FerrerSole2003}, we will use Zipf's least effort
hypothesis to derive a model within which we can frame these kind of questions properly. We will
ultimately study communication between pairs of agents sharing a given channel, so information
theory (as formulated by Claude Shannon) is the natural framework.

    In \cite{FerrerSole2003}, the tension between simplicity and communicative power proposed by Zipf
rests upon the trade-off between speaker and hearer's requirements of a language. The former prefers
to name every possible object with the same signal -- there lays their least effort to find an
object's proper name -- and the latter prefers to have a one-to-one mapping between available
signals and existing objects, so that no decoding effort is necessary. Note that the speaker's
option is the most ambiguous language possible in which communication is not possible. Meanwhile,
the hearer's proposal is not degenerated at all. The conflicting needs of different users pose an
evolutionary game for languages. These are modeled by allocations of available signals $s_i\in S$
(with $|S|=n$) to name existing objects $r_j\in R$ (with $|R|=m$). The assignments that identify a
given tongue are encoded in the entries of a matrix: $A = \{a_{ij}\}$ with $a_{ij}=1$ if signal
$s_i$ refers to object $r_j$ and $a_{ij}=0$ otherwise.

			\begin{figure}

				\includegraphics[width=11 cm]{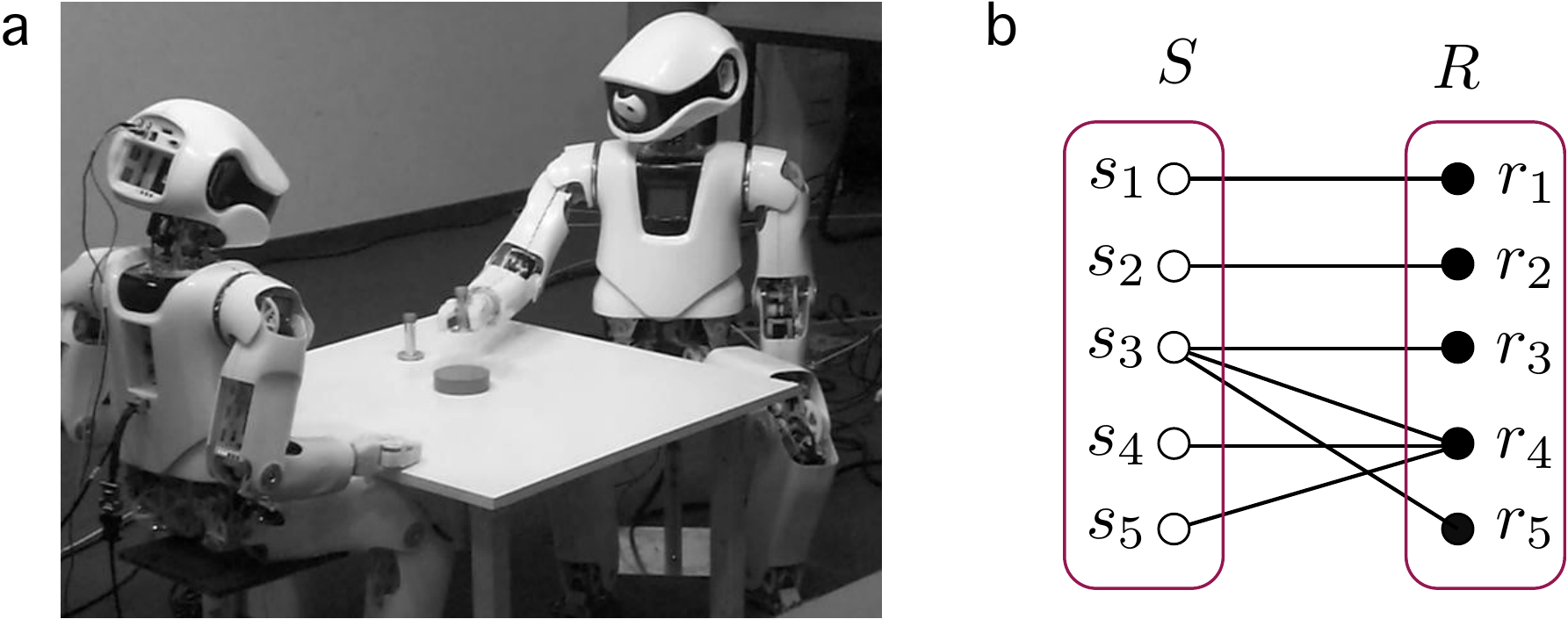}

        \caption{In order to model language evolution, one can use a number of artificial systems,
including among them robotic, embodied agents (a). Here two robots (image from the Neurocybernetics
group at Osnabr\"uck, see https://ikw.uni-osnabrueck.de/~neurokybernetik/) share a common
environment seeded by a number of objects, which they can name. Robots can evolve a rudimentary
grammar that goes beyond the simple word inventory that we could expect. Additionally, simple
mathematical models can also be used in order to capture essential features of language
organisation. A model of language can be formulated in terns of a matrix (b) that relates a set of
$n$ signals (indicated as $s_1, s_2, ..., s_n$) with a set of $m$ objects or actions of reference
($r_1, ..., r_m$). A simple case with $n=m=6$ is displayed. A signal is associated to an object
using a link connecting them. Here for example signal $s_6$ is used to refer to object $r_4$. }

				\label{fig:05.01}

			\end{figure}

    Similarly to the matrices introduced in section \ref{sec:03}, $A$ is known as an {\em adjacency
matrix}; only before it linked elements from within a set to one another and now it connects the
constituents of two different sets, $R$ and $S$, thus accounting for their relationships and other
relevant features. A very important trait is related to the presence of ambiguity. As defined, the
model and its matrix representation include both polysemy (i. e. presence of multiple meanings
associated to a given signal) as well as synonymy, where different signals refer to the same object.
The two traits can be detected by direct inspection of the rows and columns of the adjacency matrix.
If we look at the example given in figure \ref{fig:05.01}{\bf b}, using $n=m=5$ the matrix reads:
			\begin{eqnarray}
				A &=&
					\begin{pmatrix}
					 1 & 0 & 0 & 0 & 0 \\ 
					 0 & 1 & 0 & 0 & 0 \\ 
					 0 & 0 & 1 & 1 & 1 \\ 
					 0 & 0 & 0 & 1 & 0 \\ 
					 0 & 0 & 0 & 1 & 0 \\ 
					\end{pmatrix} 
					\label{eq:05.01}
			\end{eqnarray}
The A matrix structure apprehends both the capacity for a signal to have multiple meanings (by
referring to multiple objects), and synonymy, where multiple signals refer to the same object. These
two features are directly detectable here by looking at rows and columns within $A$. Synonyms are
associated to vertical strings of ones, indicating that the same object $r_k$ can be labelled or
referred to by multiple (synonymous) words. Conversely, a polysemous word would correspond to a
signal having multiple ones in a row. This contributes to the ambiguity of the language. In our
example, $r_4$ is connected to three synonyms, whereas signal $s_3$ is used to label three different
objects.

    In \cite{FerrerSole2003} it is assumed that objects are recalled randomly with uniform frequency
$p(r_i) = 1/m$. A speaker then chooses from among the available signals that name the required
object in its language $A = \{a_{ij}\}$, yielding a frequency for each signal:
			\begin{eqnarray}
				p(s_i|r_j) &=& {a_{ij} \over \omega_j}, 
				\label{eq:05.02}
			\end{eqnarray}
with $\omega_j = \sum_j a_{ij}$. We will indicate the joint probability (of having a signal and a
given object) and the corresponding  probability of a given signal as:
			\begin{eqnarray}
				p(s_i,r_j) &=& p(r_j)p(s_i|r_j), \nonumber \\
				p(s_i) &=& \sum_j p(s_i, r_j). 
				\label{eq:05.03}
			\end{eqnarray}
We can write the entropy associated to the signal diversity, which in the proposed framework stands
for the effort of the speaker:
			\begin{eqnarray}
				H_n(S) = H( \{ p(s_1), ..., p(s_n) \} ) &=& -\sum_{i=1}^n p(s_i) log_n(p(s_i)). 
				\label{eq:05.04}
			\end{eqnarray}
Recalling our needs of information theory, Shannon's entropy $H_n(S)$ provides a measure of the
underlying diversity in the system. It is also a measure of uncertainty: the higher the entropy, the
more difficult it is to predict the state of the system. For this reason $H$ is often considered a
measure of randomness. Its maximum value is obtained for a homogeneous distribution. In our case, it
corresponds to $p(s_i)=1/n$ for all signals involved:
	 		\begin{eqnarray}
				H \left ( \left\{ {1 \over n}, ..., {1 \over n} \right\} \right ) &=& 
				- \sum_{i=1}^n \left ( {1 \over n} \right ) log_n \left ( {1 \over n} \right ) = \log n. 
				\label{eq:05.05}
			\end{eqnarray}	
Conversely, the lowest entropy is obtained for $p(s_i)=1$ and $p(s_{k \ne i})=0$. For this 
single-signal scenario we obtain $H_s(S)=0$. 
		
    Another key quantity involves the noise associated to the communication channel. Using the
definition of conditional probability, namely $p(r_j | s_i) = p(s_i,r_j)/p(s_i)$ we define a measure
of noise associated to a given signal as follows:
			\begin{eqnarray}
				H_m(R|s_i) &=& -\sum_{j=1}^m p(r_j|s_i) log_m p(r_j|s_i). 
				\label{eq:05.06}
			\end{eqnarray}
This entropy weights the uncertainty associated to retrieving the right object object from $R$ when
signal $s_i$ has been used. The average uncertainty is obtained from:
			\begin{eqnarray}			 
				H_m(R|S) = \langle H_m(R|s_i) \rangle &=& \sum_{i=1}^n p(s_i) H_m(R|s_i). 
				\label{eq:05.07}
			\end{eqnarray}
For simplicity, let us assume $n=m$. If each signal were used to refer to a single and separated
object, we could order our set of objects and signals so that $p(r_j \vert s_i) = \delta_{ij}$ where
we define $\delta_{ij}=1$ for $i=j$ and zero otherwise. In this case, it is easy to see that
$H_m(R|s_i) =0$  and thus no uncertainty would be present: given a signal, the right object can be
immediately fetched without ambiguity. This corresponds to a perfect mapping between signals and
meanings/objects. The opposite case would be a completely degenerate situation where  a single
signal $s_{\mu}$ is used to refer to all objects indistinctly. Then $p(r_j \vert s_{\mu}) = 1/n$ for
all  $j=1, ..., n$. In this case, it can be shown that $H(R \vert S) = \log n$ -- thus the
uncertainty that the hearer faces is maximal.
 				
    Summing up, this conditional entropy $H(R \vert S)$ works as the average ambiguity perceived by
the hearer, and thus stands for its effort when {\em decoding} language $A=\{a_{ij}\}$. Finally,
both communicative costs are collapsed into the following energy function:
			\begin{eqnarray}
				\Omega(\lambda) &=& \lambda H_m(R|S) + (1-\lambda)H_n(S). 
				\label{eq:05.08}
			\end{eqnarray}
Using this as a kind of ``fitness" function, an evolutionary search was performed in order to
minimize $\Omega(\lambda)$. The minima obtained from this algorithm provide a picture of the
expected graphs -- as defined by the adjacency matrices -- compatible with the least effort
minimization principle.

    Along with the relative efforts defined above, two key properties were also measured. The first
is the information transfer (or mutual information) obtained from:
			\begin{eqnarray}
				I(R,S) &=& H(S) - H(S | R), 
				\label{eq:05.09}
			\end{eqnarray}
which plays a central role within information theory and is interpreted as {\em how much
information} do signals convey about which object needs to be retrieved. The second is the effective
lexicon size $\vert \cal L \vert$, i. e. the number of signals that are used to name objects. This
was defined as
			\begin{eqnarray}			 
				| {\cal L} | = \left | \left \{j | \mu_j = \sum_{k=1}^N a_{jk} > 0 \right \} \right |, 
				\label{eq:05.10}
			\end{eqnarray}
where $\mu_i$ actually indicates whether or not the signal is being used.

			\begin{figure*}
				{\centering 
					\includegraphics[width=11 cm]{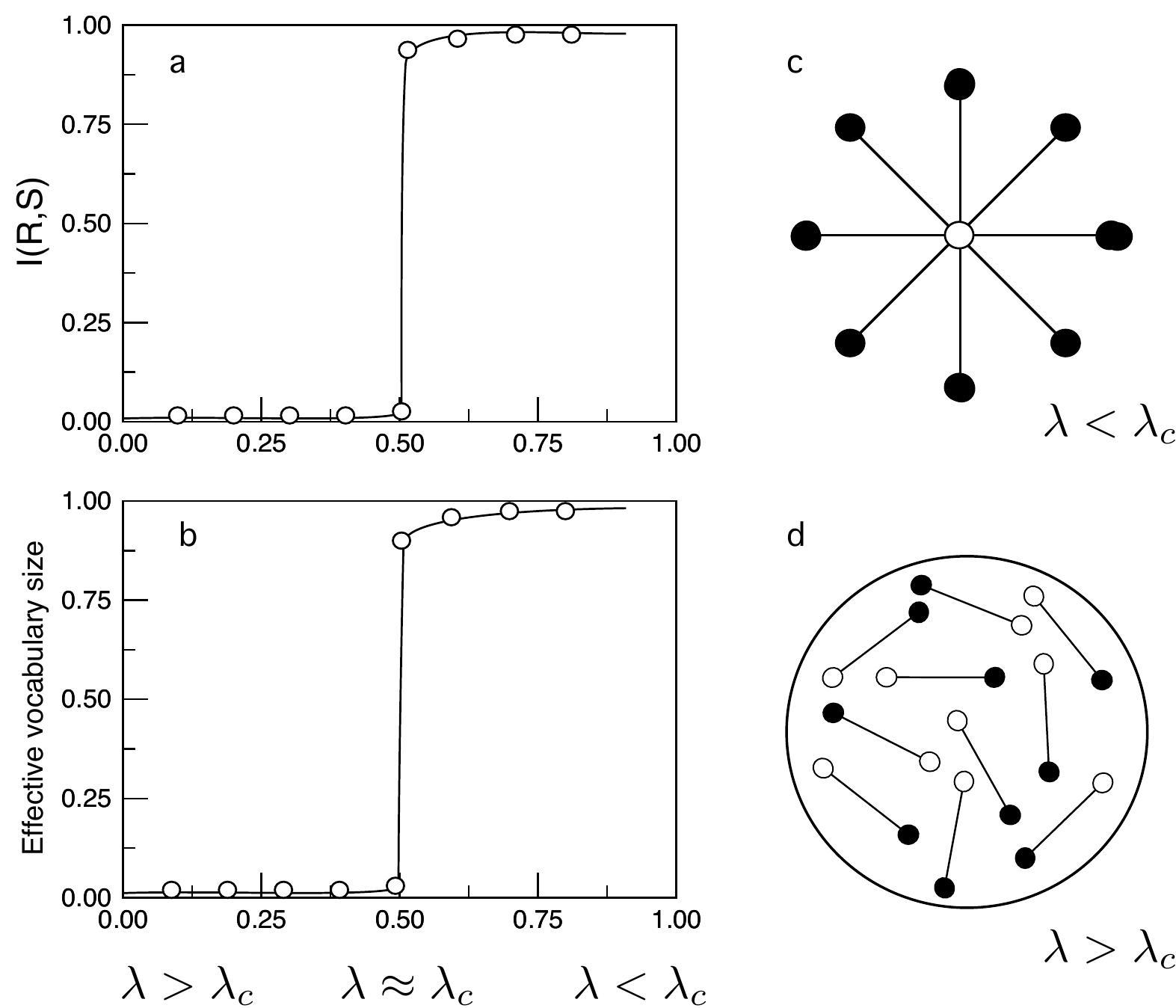}

          \caption{{\bf Phase transition in least-effort language.} As we vary $\lambda$ equation
\ref{eq:05.08} awards different importance to a speaker's or a hearer's requirements of a tongue.
Accordingly, we move from a scenario that contents the former to one that pleases the later. But the
change is sharp and happens at a very precise value of $\lambda=\lambda_c\equiv0.5$, in accordance
with the description of a first order phase transitions. The simulations to generate these plots --
top solutions after a Genetic Algorithm (GA) proceeded with different values of $\lambda$ -- are in
good agreement with this numerical critical value. Because of this sudden regime shift we can
observe very abrupt changes in some {\em order parameters} than can be measured in a language: {\bf
a} The mutual information between signals and objects (whose average value across the top population
of the GA is plotted) says how much information the signals of a language convey about the named
world. For $\lambda<\lambda_c$ one only signal serves to name every object -- fully complying with
the speaker's needs -- and the language does not bear any information about the external world, thus
communication is not feasible with such a language. For $\lambda>\lambda_c$ tongues map one-to-one
between signals and objects and a maximal amount information is conveyed. This is compared to animal
codes in \cite{FerrerSole2003}. These require a perfect mapping, thus exploit the whole range of
available signals as we can see in panel {\bf b}, where the proportion of signals used from those
available is reported. In (b) and (c) we represent the signal-object association graphs that emerge
in the two extreme regimes: $\lambda<\lambda_c$ and $\lambda>\lambda_c$ respectively. }
					
					\label{fig:05.03}
				}
			\end{figure*}

    Clearly the meta-parameter $\lambda$ weights the importance of the hearer and speaker's needs.
In \cite{FerrerSole2003} a phase transition is uncovered at a certain value $\lambda_c$ when varying
$\lambda$ between $0$ and $1$, as it is illustrated in fig. \ref{fig:05.03}. For $\lambda<\lambda_c$
the speaker's effort is minimized and completely ambiguous languages are persistently achieved. The
$A$ matrix for the extreme case in a $n=4=m$ system would be
			\begin{eqnarray}
				A =
					\begin{pmatrix}
					 1 & 1 & 1 & 1 \\
					 0 & 0 & 0 & 0 \\ 
					 0 & 0 & 0 & 0 \\ 
					 0 & 0 & 0 & 0 \\
					\end{pmatrix} 
				\label{eq:05.11}
			\end{eqnarray}
As expected, in that scenario communication is impossible, given the complete degeneracy associated
to the unique signal used to refer to every item within $R$. This is revealed by the vanishing
mutual information between signals and objects (fig. \ref{fig:05.03}{\bf a}). Obviously the
vocabulary requirements of this solution are minimal (fig. \ref{fig:05.03}{\bf b}). The word-object
association graph that we would obtain is illustrated in figure \ref{fig:05.03}{\bf c}.

    For $\lambda > \lambda_c$ the one-to-one mapping preferred by the hearer (fig.
\ref{fig:05.03}{\bf d}) is always optimal. In this special case, the adjacency matrix for the
signal-object association can be written in a diagonal form:
			\begin{eqnarray}
				A =
					\begin{pmatrix}
					 1 & 0 & 0 & 0 \\
					 0 & 1 & 0 & 0 \\ 
					 0 & 0 & 1 & 0 \\ 
					 0 & 0 & 0 & 1 \\
					\end{pmatrix}. 
				\label{eq:05.12}
			\end{eqnarray}
Most models of language evolution that explore the origins of communication under natural selection
end up in finding these type of diagonal matrices.This is
compared to animal communicative systems in \cite{FerrerSole2003}. Such systems present  a non-
degenerated mapping between objects and signals. The exhaustive vocabulary needs of this regime is
illustrated in figure \ref{fig:05.03}{\bf b}. This case would be favored in a scenario where few
signals suffice for communication, and it would be restrained by the memory capacities of hearer and
speaker. Indeed, it has been shown how memory constrains could prompt the development of a fully
articulated human language when vocabulary size overcomes a certain threshold \cite{NowakJansen2000}
so that units might be reused, but at the expense of making them ambiguous. In the least-effort
framework proposed in \cite{FerrerSole2003}, such a language would show up only at the phase
transition $\lambda\sim\lambda_c$. Then, hearer and speaker's needs are equally taken into account,
language instances with a moderate level of ambiguity are found, and communication is still possible
-- as the sharply varying mutual information between signals and objects around $\lambda_c$ points
out (fig. \ref{fig:05.03}{\bf a}).

    In the original work \cite{FerrerSole2003} the phase transition reported was of second order, meaning
that the shift from non-communicative codes to one-to-one mappings was a smooth drift across several
intermediate steps -- any of them could be a relatively fit candidate of human language, not so
urgently needing to tune $\lambda$ to its critical value $\lambda_c$. But further investigation of
the problem clearly indicates that the transition is of first order in nature and that $\lambda_c =
0.5$ (for m=n), as figure \ref{fig:05.03} clearly shows. This means that the jump between the two
extreme cases happens swiftly at $\lambda=0.5$, that a graduated range of possibilities that solve
the optimization problem for $\lambda \sim \lambda_c$ does not exist, and that only at $\lambda =
\lambda_c$ could we find a phenomenology akin to human language.

    The analysis in \cite{FerrerSole2003} is complemented with an investigation of the frequency with
which different signals show up for a given language in the model. This can be made thanks to
equations \ref{eq:05.02} and \ref{eq:05.03}. Remarkably, at the phase transition it was found that
the frequency of different words obey Zipf's law \cite{FerrerSole2003}, thus closing the circle with one
of the observations that opened our quest.\\

    This work \cite{FerrerSole2003} has been featured here for its historical importance in
promoting the least-effort language agenda. However, its results have been contested and can not be
held as correct anymore without a critical revision. The first and foremost claim has been that the
algorithm employed in \cite{FerrerSole2003} usually only achieves local minimization, thus the
portrayed languages would not correspond to global least-effort codes \cite{ProkopenkoPolani2010}.
Furthermore, when analyzing the global optima of the problem we find ourselves with a degenerated
solution -- i.e. multiple assignments between objects and signals optimize the trade-off between
speaker and hearer needs at the phase transition \cite{Ferrer2005b, FerrerDiaz2007,
ProkopenkoPolani2010}.

   Three observations are pertinent about these critics: i) Among the several solutions to the
least-effort problem at the indicated phase transitions we find Zipf's law as well
\cite{ProkopenkoPolani2010}. This is not the dominating solution, though -- i.e. there are more
solutions with some other frequency distribution of signals than solutions whose signal usage
follows equations \ref{eq:02.01} and \ref{eq:02.02} \cite{ProkopenkoPolani2010}. Thus we would
expect that when choosing randomly among all least-effort solutions for $\lambda=\lambda_c$ we would
likely arrive to some other distribution but to Zipf's. However, ii) the original investigation of
least-effort communicative systems and the framework that this model introduces remain valid and
very appealing, even if they do not suffice to produce Zipf's law. The least-effort principle has
still got robust experimental and theoretical motivations, and we should not discard further forces
operating upon language evolution that would select Zipf's law against others. In such a case, the
least-effort game described in this section would be just a sub-problem that language evolution has
solved over time. Finally, iii) concerning the main topic of this volume; even if Zipf's law were
not recovered, robust evidence exists indicating that the trade-off posed by the least-effort
procedure is a way in of ambiguity into human language.

    The featured model has been furthered by successive works. The hunt for a robust mechanism that
generates the Zipf distribution continues and interesting proposals are being explored. A very
promising one relies on the open-ended nature of human language \cite{CorominasSole2011}. Previous
work by the same authors showed how Zipf's law is unavoidable in a series of stochastic systems. A
key feature of those systems is that they grow by sampling an infinite number of states
\cite{CorominasSole2010}. When applied to language, not only the unboundedness of human language is
necessary but also the sempiternal least effort, so that Zipf's law can be successfully obtained for
communicating systems. Interestingly, the approach in \cite{CorominasSole2011} applies the 
least-effort principle upon the transition between stages of the language as it grows in size -- by
incorporating new signals to its repertoire. This explicit role of the contingent historical path is
an interesting lead absent in the main body of literature. A slightly different research line
followed by these authors uses the proposed model to quantify precisely how much information is lost
due to the ambiguity of optimal languages when the trade-offs discussed above are satisfied
\cite{FortunyCorominas2013, CorominasSole2014}.

    Finally, several authors elaborate upon the model described above. In the critical review
mentioned earlier \cite{ProkopenkoPolani2010} it is noted how the original model is not sufficient
to always derive Zipf's law for the optimal model languages. The authors modify equation
\ref{eq:05.08} and propose:
			\begin{eqnarray}
				\Omega(\lambda)^0 &=& -\lambda I(S|R) + (1-\lambda)H(S) = -\lambda H(R) + \Omega(\lambda), 
				\label{eq:05.13}
			\end{eqnarray}
as a target for minimization; where $I(S|R)$ is the mutual information between signals and objects
in the sets $S$ and $R$ respectively. This new target becomes eq. \ref{eq:05.08} if all objects are
equally probable. Equation \ref{eq:05.13} is more adequate to ``better account for subtle
communication efforts'' \cite{ProkopenkoPolani2010}, as more costs implicit in equation
\ref{eq:05.13} but absent in equation \ref{eq:05.08} are considered. In a follow up paper
\cite{SalgeProkopenko2013} it is demonstrated how an ingredient to robustly derive Zipf's law in
their model is to take into account signal costs, which makes sense considering that different
signals require different time, effort, or energy to be produced, broadcast, collected, and
interpreted. This, as we will see in the following section, can also be an important element for the
presence of ambiguity in human languages.

	\section{Ambiguity, principles of information theory, and least effort}
		\label{sec:06}

    Several recent empirical observations illustrate an optimization force -- that justifies our
least effort point of view -- acting upon different linguistic facets such as prosody, syntax,
phonology, and many others \cite{LevyJaeger2007, FrankJaeger2008, Jaeger2010, PiantadosiGibson2011a,
MahowaldGibson2013}. This evidence accumulates with other, previously shown global-level language
organizational features epitomized by the properties of the small worlds (sects. \ref{sec:03} and
\ref{sec:04}). All this indicates that optimization principles and natural selection should play a
paramount role to understanding human communication in a broad sense. As we have seen, entropies
arise or need to be explicitly introduced with a twofold purpose: as a metric and as a specific
optimization target. The ubiquity of this mathematical construct -- that, we recall, gives a measure
of degeneracy and, more specifically in our context, of degeneracy of meanings -- is a first clue
that the price to pay for a least effort language is ambiguity, as we will argue right below again
and as suggested by the results from section \ref{sec:05}.

    In \cite{PiantadosiGibson2011b} a formalization of this trade-off between least-effort and
ambiguity is presented. They argue that any optimal code will be ambiguous when examined out of
context, provided the context offers redundant information; and they do so presenting extremely
elegant, easy, and powerful information theoretical arguments that apply beyond human communication.
Specially the first argument is of general validity for {\em any communicative system} within a
context that is informative about a message. The two alternative -- but similar -- paths that the
authors provide towards ambiguity are the following (the quotes are from
\cite{PiantadosiGibson2011b}):
			\begin{itemize}

        \item {\em ``Where context is informative about meaning, unambiguous language is partly
redundant with the context and therefore inefficient.''}

          The authors conceive a space $M$ consisting of all possible meanings $m$ such that
inferring a precise meaning out of a signal demands at least
						\begin{eqnarray}
							H[M] &=& -\sum_{m\in M}P(m)log\{P(m)\}
							\label{eq:06.01}
						\end{eqnarray}
bits of information, with $P(m)$ the probability that meaning $m$ needs to be recalled. Similarly,
they assume a space $C$ that encompasses all possible contexts $c$, compute the entropy of each
meaning conditioned to happen within each context, and average over contexts:
						\begin{eqnarray}
							H[M|C] &=& \sum_{c\in C} P(c) \sum_{m\in M} P(m|c)log\{P(m|c)\}. 
							\label{eq:06.02}
						\end{eqnarray}
This accounts for the average number of information (in bits) that a code needs to provide to tell
apart different meanings within discriminative enough contexts. If context is informative it is
likely that $H[M]>H[M|C]$ \cite{PiantadosiGibson2011b, CoverThomas2012}.

          With this in hand the authors have shown how ``the least amount of information that a
language can convey without being ambiguous is $H[M|C]$'', which is lower than $H[M]$; thus any
optimal code will seem ambiguous when examined out of context and any unambiguous code will be
suboptimal in that it produces more information than strictly necessary.

          Note once more the elegance of the argument and its generality: no requirements are made
about the meanings or the contexts, and the later are general enough as to include any circumstance
of any kind affecting communication in any way.

      	\item {\em ``Ambiguity allows the re-use of words and sounds which are more easily produced or
understood.''}

          This second argument only diminishes in generality because the authors must consider that
different signals in a code vary in cost -- i.e. that they are not of equal length or complexity, or
that distinct signs require different amount of effort when they are used. This becomes obvious in
human speech, e.g., considering the longer time that larger words demand. Note anyway that this is a
quite general scenario still affecting most conceivable communicative systems and, of course, any
kind of human communication.

          The argument acknowledges that it is preferable to use simpler signals. Then, ambiguity
enables us to re-use the same signal in different contexts, assuming always that the context
provides the needed disambiguation. 

			\end{itemize}

    According to these ideas, that optimal codes are ambiguous if the context is informative does
not imply that human languages must be ambiguous, neither that any ambiguous coding is more optimal
than any unambiguous one. However, ambiguity -- say polysemy, in certain contexts, but not only --
is an extremely extended phenomenon in human language when tongues are analyzed out of context, and
the authors propose that such simple yet forceful reasoning explains its pervasiveness. In previous
sections a much stronger point was made based on empirical observations: this polysemy not only does
exist, but it also shapes the structure of tongues such that a global order arises in many aspects
of it (e.g. semantic networks), and such that it presents very convenient features that render human
language optimal or very effective (e.g. for semantic navigation). Thus not only ambiguity is
present, it seems to be of a very precise kind in order to comply with several optimization needs at
a same time, such as Zipf's least effort paradigm proposed \cite{Zipf1949}.\\

	\section{Discussion and prospects}
		\label{sec:07}

    The models and real networks presented above provide a well-defined theoretical and quantitative
framework to address language structure and its evolution. The sharp transition between 
non-communicative and communicative phases is a remarkable finding -- and the fact that intuitive
models can reproduce this feature is impressive. This suggests that a fundamental property
associated to the least effort minimization principle involves an inevitable gap to be found among
its solutions. From another perspective, both real language networks and the simple graphs emerging
from the least effort algorithm(s) introduce ambiguity as a natural outcome of their heterogeneous
nature.

    While the path explored this far invites us to be optimistic, several open problems arise from
the results reviewed. These will require further research until a complete picture of human language
-- beyond the role of ambiguity -- is attained. Here is a tentative list of open issues:
			\begin{enumerate}

        \item Both the topological analysis of semantic networks and what can be proposed from
simple models are typically disconnected from an explicit cognitive substrate. Some remarkable works
on semantic webs have shown that the structure of semantic webs includes a modular organization
where groups of semantically related words are more connected among them than with other items.
Individuals mentally searching on this space seem to make fast associations between items within
modules as well as seemingly random jumps between modules \cite{GoniVilloslada2011}. The pattern of
search is actually related to the ways search is performed on computer networks. Moreover, there is
a literature on neural network models of semantic association \cite{MartinChao2001, HuthGallant2012}
that could be explored in order to see how the space of neural attractors and the underlying
categorization emerging from them are linked to a semantic network. Models of damage in semantic
webs (using topological methods) already suggest that relevant information might be obtained in
relation with the process of cognitive decay associated to some neurodegenerative diseases
\cite{ChertkowSeidenberg1989, BorgeArenas2011}.

        \item A very promising field within language evolution involves using embodied agents
(robots or physical simulations of them) that are capable of learning, memory, and association
\cite{Steels2003}. A protogrammar has been shown to emerge in these embodied communicating agents
\cite{Steels2000, Steels2012, BeulsSteels2013}. The study of lexical and grammatical processing in
these robotic agents using so called Fluid Construction Grammars (FCGs) \cite{SteelsDeBeule2006,
Steels2011} reveals that language evolution might take place by optimizing lexicon size and the
construction structures in order to minimize search. More traditional approaches to computer
languages -- as in programming languages -- explicitly reject ambiguity for the challenges it
presents. It is made clear that FCGs seek more malleable structures (thus the {\em Fluid}), ready to
evolve and be adopted and adapted by a population -- in this case, of robots. The population is
usually not expected to share the exact same grammatic structure as it emerges, thus clearing a path
for ambiguity. Notwithstanding this, part of the problems solved by this novel approach is one of
reducing ambiguity out of the messages being interchanged by the talking agents
\cite{SteelsNeubauer2005}. Also, the emergence of grammatical rules is a direct consequence of this
ambiguity reduction \cite{BeulsSteels2013}.

        \item In all studies so far developed, models of language evolution involve only one type
of network level of description. However, semantic, syntactic and even phonologic levels interact
and any relevant analysis should include several network levels. How are different networks connected 
to each other? What is the impact of their special topological and scaling properties on the 
global behavior of language as a whole? 

        \item Statistical physics is at the core of many of the approximations considered in this
paper. Despite the biological nature of language and its historical origins, we have seen that some
strong regularities are inevitable and are more fundamental than we would expect. There are many
other ways of approaching language structure using these methods, including the analysis of language
ontogeny \cite{CorominasSole2009, BaixeriesFerrer2013} and the structure of syntactic networks.
Available evidence from data and models suggests that, once Zipf's law is at work, a number of well
known regularities exhibited by syntax graphs are obtained \cite{FerrerBollobas2005}. This would be
consistent with an evolutionary scenario where syntax might come for free, as a byproduct of
possibly inevitable features of correlations among words following Zipf's law \cite{Sole2005}. The
idea is appealing and worth researching and, once again, complex networks and information theory
might provide a valid framework.

        \item A twin problem to that of ambiguity is revealed when we consider synonymy. This trait
might be a contingency, and it is considered rare by scholars \cite{NowakKrakauer1999}. Indeed,
while different models account for it \cite{NowakKrakauer1999, FerrerSole2003,
SalgeProkopenko2013}, all of them predict that synonymy should not be present in optimal languages
or languages in equilibrium; but yet we observe some degree of synonymy in every human code. 

			\end{enumerate}


\begin{thebibliography}{99}

	\bibitem[Albert and Barabasi 2002]{AlbertBarabasi2002}
		Albert, R\'eka, and Albert-L\'aszl\'o Barab\'asi. 
		``Statistical mechanics of complex networks.'' 
		Reviews	of modern physics 74, no. 1 (2002): 47.

	\bibitem[Altmann et al. 2012]{AltmannEsposti2012}
		Altmann, Eduardo G., Giampaolo Cristadoro, and Mirko Degli Esposti. 
		``On the origin of long-range correlations in texts.''
		Proceedings of the National Academy of Sciences 109, no. 29 (2012): 11582-11587.

	\bibitem[Baixeries et al. 2013]{BaixeriesFerrer2013}
		Baixeries, Jaume, Brita Elvev\r{a}g, and Ramon Ferrer-i-Cancho. 
		``The Evolution of the Exponent of Zipf's Law in Language Ontogeny.'' 
		PloS one 8, no. 3 (2013): e53227.

	\bibitem[Beuls and Steels 2013]{BeulsSteels2013}
		Beuls, Katrien, and Luc Steels. 
		``Agent-based models of strategies for the emergence and evolution of grammatical agreement.'' 
		PloS one 8, no. 3 (2013): e58960.

	\bibitem[Borge-Holthoefer et al. 2011]{BorgeArenas2011}
		Borge-Holthoefer, Javier, Yamir Moreno, and Alex Arenas. 
		``Modeling abnormal priming in Alzheimer's patients with a free association network.'' 
		PloS one 6, no. 8 (2011): e22651.

	\bibitem[Chan et al. 1997]{ChanSalmon1997}
		Chan, Agnes S., Nelson Butters, and David P. Salmon. 
		``The deterioration of semantic networks in patients with Alzheimer's disease: A cross-sectional study.'' 
		Neuropsychologia 35, no. 3 (1997):241-248.

	\bibitem[Chertkow et al. 1989]{ChertkowSeidenberg1989}
		Chertkow, Howard, Daniel Bub, and Mark Seidenberg. 
		``Priming and semantic memory loss in Alzheimer's disease.'' 
		Brain and language 36, no. 3 (1989): 420-446.

	\bibitem[Chomsky 2000]{Chomsky2000}
		Chomsky, Noam. 
		Minimalist inquiries: The framework. No. 15. 
		MIT Working Papers in Linguistics, MIT, Department of Linguistics, 1998.

	\bibitem[Chomsky 2002]{Chomsky2002}
		Chomsky, Noam. 
		``An interview on minimalism.'' 
		N. Chomsky, On Nature and Language (2002): 92-161.

	\bibitem[Corominas-Murtra et al. 2009]{CorominasSole2009}
		Corominas-Murtra, Bernat, Sergi Valverde, and Ricard Sole. 
		``The ontogeny of scale-free syntax networks: phase transitions in early language acquisition.'' 
		Advances in Complex Systems 12, no. 03 (2009): 371-392.

	\bibitem[Corominas-Murtra and Sol\'e 2010]{CorominasSole2010}
		Corominas-Murtra, Bernat, and Ricard V. Sol\'e. 
		``Universality of Zipf's law.'' 
		Physical Review E 82, no. 1 (2010): 011102.

	\bibitem[Corominas-Murtra and Sol\'e 2011]{CorominasSole2011}
		Corominas-Murtra, Bernat, Jordi Fortuny, and Ricard V. Sol\'e. 
		``Emergence of Zipf's law in the evolution of communication.'' 
		Physical Review E 83, no. 3 (2011): 036115.

	\bibitem[Corominas-Murtra et al. 2014]{CorominasSole2014}
		Corominas-Murtra, Bernat, Jordi Fortuny, and Ricard V. Sol\'e
		``Towards a mathematical theory of meaningful communication.''
		arXiv pre-print: http://arxiv.org/abs/1004.1999 (2014). 

	\bibitem[Cover and Thomas 2012]{CoverThomas2012}
		Cover, Thomas M., and Joy A. Thomas. 
		Elements of information theory. 
		John Wiley \& Sons, 2012.

	\bibitem[Fellbaum 1998]{Fellbaum1998}
		Fellbaum, Christiane, ed. 
		WordNet: An Electronic Lexical Database. 
		Cambridge, MA: MIT Press, 1998. 

		Steels, Luc, ed. 
		Design patterns in fluid construction grammar. Vol. 11. 
		John Benjamins Publishing, 2011.

	\bibitem[Ferrer i Cancho and Sol\'e 2001a]{FerrerSole2001a}
		Ferrer i Cancho, Ramon, and Ricard V. Sol\'e. 
		``The small world of human language.'' 
		Proceedings of the Royal Society of London. Series B: Biological Sciences 268, no. 1482 (2001): 2261-2265.

	\bibitem[Ferrer i Cancho and Sol\'e 2001b]{FerrerSole2001b}
		Ferrer i Cancho, Ramon, and Ricard V. Sol\'e. 
		``Two Regimes in the Frequency of Words and the Origins of Complex Lexicons: Zipf's Law Revisited." 
		Journal of Quantitative Linguistics 8, no. 3 (2001): 165-173.

	\bibitem[Ferrer i Cancho and Sol\'e 2001c]{FerrerSole2001c}
		Ferrer i Cancho, Ramon, and Ricard V. Sol\'e. 
		``Zipf's law and random texts.'' 
		Advances in Complex Systems 5, no. 01 (2002): 1-6.

	\bibitem[Ferrer i Cancho and Sol\'e 2003]{FerrerSole2003}
		Ferrer i Cancho, Ramon, and Ricard V. Sol\'e. 
		``Least effort and the origins of scaling in human language.'' 
		Proceedings of the National Academy of Sciences 100, no. 3 (2003): 788-791.

	\bibitem[Ferrer i Cancho et al. 2004]{FerrerKohler2004}
		Ferrer i Cancho, Ramon, Ricard V. Sol\'e, and Reinhard K\"ohler. 
		``Patterns in syntactic dependency networks.'' 
		Physical Review E 69, no. 5 (2004): 051915.

	\bibitem[Ferrer i Cancho 2005a]{Ferrer2005a}
		Ferrer i Cancho, Ramon. 
		``The variation of Zipf's law in human language.'' 
		The European Physical Journal B-Condensed Matter and Complex Systems 44, no. 2 (2005): 249-257.

	\bibitem[Ferrer i Cancho 2005b]{Ferrer2005b}
		Ferrer i Cancho, Ramon. 
		``Decoding least effort and scaling in signal frequency distributions.'' 
		Physica A: Statistical Mechanics and its Applications 345, no. 1 (2005): 275-284.

	\bibitem[Ferrer i Cancho et al. 2005]{FerrerBollobas2005}
		Ferrer i Cancho, Ramon, Oliver Riordan, and B\'ela Bollob\'as. 
		``The consequences of Zipf's law for syntax and symbolic reference.'' 
		Proceedings of the Royal Society B: Biological Sciences 272, no. 1562 (2005): 561-565.

	\bibitem[Ferrer i Cancho and D\'iaz-Guilera 2007]{FerrerDiaz2007}
		Ferrer i Cancho, Ramon, and Albert D\'iaz-Guilera. 
		``The global minima of the communicative energy of natural communication systems.'' 
		Journal of Statistical Mechanics: Theory and Experiment 2007, no. 06 (2007): P06009.

	\bibitem[Frank and Jaeger 2008]{FrankJaeger2008}
		Frank, Austin, and Tim F. Jaeger. 
		``Speaking rationally: Uniform information density as an optimal strategy for language production.'' 
		CogSci. Washington, DC: CogSci (2008).

	\bibitem[Fortuni and Corominas-Murtra 2013]{FortunyCorominas2013}	
		Fortuny, Jordi, and Bernat Corominas-Murtra. 
		``On the origin of ambiguity in efficient communication.'' 
		Journal of Logic, Language and Information 22, no. 3 (2013): 249-267.

	\bibitem[Go\~ni et al. 2011]{GoniVilloslada2011}
		Go\~ni, Joaqu\'in, Gonzalo Arrondo, Jorge Sepulcre, I\~nigo Martincorena, V\'elez de Mendiz\'abal,
		Nieves, Corominas-Murtra, Bernat, Bejarano, Bartolom\'e et al.
		``The semantic organization of the animal category: evidence from semantic verbal fluency and network theory.'' 
		Cognitive processing 12, no. 2 (2011): 183-196.

	\bibitem[Gregory 2008]{Gregory2008}
		Gregory, T. Ryan. 
		``The evolution of complex organs.'' 
		Evolution: Education and Outreach 1, no. 4 (2008): 358-389.

	\bibitem[Hauser et al. 2002]{HauserFitch2002}
		Hauser, Marc D., Noam Chomsky, and W. Tecumseh Fitch. 
		``The faculty of language: What is it, who has it, and how did it evolve?.'' 
		Science 298, no. 5598 (2002): 1569-1579.

	\bibitem[Holanda et al. 2004]{HolandaSeron2004}
		Holanda, Adriano de Jesus, Ivan Torres, Osame Kinouchi, Alexandre Souto, and Evandro E. Seron. 
		``Thesaurus as a complex network.'' 
		Physica A: Statistical Mechanics and its Applications 344, no. 3 (2004): 530-536.

	\bibitem[Huth et al. 2012]{HuthGallant2012}
		Huth, Alexander G., Shinji Nishimoto, An T. Vu, and Jack L. Gallant. 
		``A continuous semantic space describes the representation of thousands of object and action categories across the human brain.'' 
		Neuron 76, no. 6 (2012): 1210-1224.

	\bibitem[Jaeger 2010]{Jaeger2010}
		Jaeger, Tim F. 
		``Redundancy and reduction: Speakers manage syntactic information density.'' 
		Cognitive psychology 61, no. 1 (2010): 23-62.

	\bibitem[Kauffman 1993]{Kaufman1993}
		Kauffman, Stuart A. 
		The origins of order: Self-organization and selection in evolution. 
		Oxford university press, 1993.

	\bibitem[Ke 2004]{Ke2004}
		Ke, Jinyun. 
		``Self-organization and language evolution: system, population and individual.'' 
		PhD diss., City University of Hong Kong, 2004.

	\bibitem[Kinouchi et al. 2002]{KinouchiRisau2002}
		Kinouchi, Osame, Alexandre S. Martinez, Gilson F. Lima, G. M. Lourenço, and Sebastian Risau-Gusman. 
		``Deterministic walks in random networks: An application to thesaurus graphs.'' 
		Physica A: Statistical Mechanics and its Applications 315, no. 3 (2002): 665-676.

	\bibitem[Levy and Jaeger 2007]{LevyJaeger2007}
		Levy, Roger, and Tim F. Jaeger. 
		``Speakers optimize information density through syntactic reduction.'' 
		Advances in neural information processing systems 19 (2007): 849.

	\bibitem[Mahowald et al. 2013]{MahowaldGibson2013}
		Mahowald, Kyle, Evelina Fedorenko, Steven T. Piantadosi, and Edward T. Gibson. 
		``Speakers choose shorter words in predictive contexts.''
		Cognition 126 no. 2 (2013): 313-318. 

	\bibitem[Mantegna et al. 1991]{MantegnaStanley1994}
		Mantegna, R. N., S. V. Buldyrev, A. L. Goldberger,S. Havlin, C. K. Peng, M. Simons, and H. E. Stanley. 
		``Linguistic features of noncoding DNA sequences.'' 
		Physical review letters 73, no. 23 (1994): 3169.

	\bibitem[Martin and Chao 2001]{MartinChao2001}
		Martin, Alex, and Linda L. Chao. 
		``Semantic memory and the brain: structure and processes.'' 
		Current opinion in neurobiology 11, no. 2 (2001): 194-201.

	\bibitem[Maynard-Smith and Szathm\'ary 1995]{MaynardSzathmary1995}
		Smith, John Maynard, and E\"ors Szathm\'ary. 
		The major transitions in evolution. 
		Oxford University Press, 1997.

	\bibitem[Milgram 1967]{Milgram1967}
		Milgram, Stanley. 
		``The small world problem.'' 
		Psychology today 2, no. 1 (1967): 60-67.

	\bibitem[Miller 1995]{Miller1995}
		Miller, George A. 
		``WordNet: A Lexical Database for English. ''
		Communications of the ACM Vol. 38, No. 11 (1995): 39-41. 

	\bibitem[Motter et al. 2002]{MotterDasgupta2002}
		Motter, Adilson E., Alessandro P. S. de Moura, Ying-Cheng Lai, and Partha Dasgupta. 
		``Topology of the conceptual network of language.'' 
		Physical Review E 65, no. 6 (2002): 065102.

	\bibitem[Nowak et al. 1999]{NowakKrakauer1999}
		Nowak, Martin A., Joshua B. Plotkin, and David C. Krakauer. 
		``The evolutionary language game." 
		Journal of Theoretical Biology 200, no. 2 (1999): 147-162.

	\bibitem[Nowak et al. 2000]{NowakJansen2000}
		Nowak, Martin A., Joshua B. Plotkin, and Vincent AA Jansen. 
		``The evolution of syntactic communication.'' 
		Nature 404, no. 6777 (2000): 495-498.

	\bibitem[Obst et al. 2011]{ObstProkopenko2011}
		Obst, Oliver, Daniel Polani, and Mikhail Prokopenko. 
		``Origins of scaling in genetic code.'' 
		In Advances in Artificial Life. Darwin Meets von Neumann, pp. 85-93. Springer Berlin Heidelberg, 2011.

	\bibitem[Peterson et al. 2012]{PetersenPerc2012}
		Petersen, Alexander M., Joel N. Tenenbaum, Shlomo Havlin, H. Eugene Stanley, and Matjaž Perc. 
		``Languages cool as they expand: Allometric scaling and the decreasing need for new words.'' 
		Scientific reports 2 (2012).

	\bibitem[Piantadosi and Gibson 2011]{PiantadosiGibson2011a}
		Piantadosi, Steven T., Harry Tily, and Edward Gibson. 
		``Word lengths are optimized for efficient communication.''
		Proceedings of the National Academy of Sciences 108, no. 9 (2011): 3526-3529.

	\bibitem[Piantadosi et al. 2011]{PiantadosiGibson2011b}
		Piantadosi, Steven T., Harry Tily, and Edward Gibson. 
		``The communicative function of ambiguity in language.'' 
		Cognition 122, no. 3 (2012): 280-291.

	\bibitem[Pinker and Bloom 1990]{PinkerBloom1990}
		Pinker, Steven, and Paul Bloom. 
		``Natural language and natural selection.'' 
		Behavioral and brain sciences 13, no. 04 (1990): 707-727.

	\bibitem[Prokopenko et al. 2010]{ProkopenkoPolani2010}
		Polani, Daniel. 
		``Phase transitions in least-effort communications.'' 
		Journal of Statistical Mechanics: Theory and Experiment (2010).

	\bibitem[Pustejovsky 1991]{Pustejovsky1991}
		Pustejovsky, James. 
		``The generative lexicon.'' 
		Computational linguistics 17, no. 4 (1991): 409-441.

	\bibitem[Pustejovsky 1995]{Pustejovsky1995}
		Pustejovsky, James. 
		The generative lexicon. 
		MIT Press, Cambridge, MA, 1995. 

	\bibitem[Salge et al. 2013]{SalgeProkopenko2013}
		Salge, Christoph, Nihat Ay, Daniel Polani, and Mikhail Prokopenko. 
		``Zipf's Law: Balancing Signal Usage Cost and Communication Efficiency.'' 
		SFI working paper: 13-10-033 (2013).

	\bibitem[Scalise 1984]{Scalise1984}
		Scalise, Sergio. 
		Generative morphology. Vol. 18. 
		Walter de Gruyter, 1986.

	\bibitem[Searls 2002]{Searls2002}
		Searls, David B. 
		``The language of genes.'' 
		Nature 420, no. 6912 (2002): 211-217.

	\bibitem[Sigman and Cecchi 2002]{SigmanCecchi2002}
		Sigman, Mariano, and Guillermo A. Cecchi. 
		``Global organization of the Wordnet lexicon.'' 
		Proceedings of the National Academy of Sciences 99, no. 3 (2002): 1742-1747.

	\bibitem[Sol\'e and Goodwin 2001]{SoleGoodwin2001}
		Sole, Ricard, and Brian Goodwin. 
		Signs of life: How complexity pervades biology. 
		Basic books, 2008.

	\bibitem[Sol\'e 2005]{Sole2005}
		Sol\'e, Ricard. 
		``Language: Syntax for free?.'' 
		Nature 434, no. 7031 (2005): 289-289.

	\bibitem[Sol\'e et al. 2010]{SoleFortuny2010}
		Sol\'e, Ricard V., Bernat Corominas-Murtra, and Jordi Fortuny. 
		``Diversity, competition, extinction: the ecophysics of language change.'' 
		Journal of The Royal Society Interface 7, no. 53 (2010): 1647-1664.

	\bibitem[Steels 2000]{Steels2000}
		Steels, Luc. 
		``The emergence of grammar in communicating autonomous robotic agents.'' 
		In ECAI, pp. 764-769. 2000.

	\bibitem[Steels 2003]{Steels2003}
		Steels, Luc. 
		``Evolving grounded communication for robots.'' 
		Trends in cognitive sciences 7, no. 7 (2003): 308-312.

	\bibitem[Steels et al. 2005]{SteelsNeubauer2005}
		Steels, Luc, Joachim De Beule, and Nicolas Neubauer. 
		``Linking in Fluid Construction Grammars.'' 
		In BNAIC, pp. 11-20. 2005.

	\bibitem[Steels and De Beule 2006]{SteelsDeBeule2006}
		Steels, Luc, and Joachim De Beule. 
		``A (very) brief introduction to fluid construction grammar.'' 
		In Proceedings of the Third Workshop on Scalable Natural Language Understanding, pp. 73-80. Association for Computational Linguistics, 2006.

	\bibitem[Steels ed. 2011]{Steels2011}
		Steels, Luc, ed. 
		Design patterns in fluid construction grammar. Vol. 11. 
		John Benjamins Publishing, 2011.

	\bibitem[Steels ed. 2012]{Steels2012}
		Steels, Luc, ed. 
		Experiments in cultural language evolution. Vol. 3. 
		John Benjamins Publishing, 2012.

	\bibitem[Steyvers and Tenenbaum 2005]{SteyversTenenbaum2005}
		Steyvers, Mark, and Joshua B. Tenenbaum. 
		``The Large-Scale Structure of Semantic Networks: Statistical Analyses and a Model of Semantic Growth.'' 
		Cognitive science 29, no. 1 (2005): 41-78.

	\bibitem[Trubetskoi 1939]{Trubetskoi1939}
		Trubetzkoy, Nikolai S. 
		Principles of Phonology. 
		Reprinted 1969, University of California Press, Berkeley, CA. 

	\bibitem[Valverde et al. 2002]{ValverdeSole2002}
		Valverde, Sergi, R. Ferrer Cancho, and Richard V. Sole. 
		``Scale-free networks from optimal design.'' 
		EPL (Europhysics Letters) 60, no. 4 (2002): 512.

	\bibitem[Watts and Strogatz 1998]{WattsStrogatz1998}
		Watts, Duncan J., and Steven H. Strogatz. 
		``Collective dynamics of `small-world' networks.'' 
		Nature 393, no. 6684 (1998): 440-442.

	\bibitem[Wray ed. 2002]{Wray2002}
		Wray, Alison, ed. 
		The transition to language. Vol. 2. 
		Peterson's, 2002.

	\bibitem[Zipf 1949]{Zipf1949}
		Zipf, George Kingsley. 
		``Human behavior and the principle of least effort.'' (1949).


\end{thebibliography}
\end{document}